\documentclass[11pt,a4paper]{article}
\usepackage[left=20mm, top=20mm, right=20mm, bottom=20mm, nohead=true, nofoot=true]{geometry}
\usepackage[utf8]{inputenc}
\usepackage{authblk}
\usepackage{indentfirst}
\usepackage{misccorr}
\usepackage{graphicx}
\usepackage{amsmath}
\usepackage{amssymb}
\usepackage{multicol}
\usepackage{bm}
\usepackage{longtable}
\usepackage{pdflscape}
\usepackage{epstopdf}
\usepackage{multirow}
\usepackage{adjustbox}
\usepackage{amsfonts}
\usepackage{ifthen}
\usepackage{fancyhdr}
\usepackage{setspace}
\usepackage{xcolor}
\usepackage[round,authoryear,comma,sort&compress,numbers]{natbib}
\usepackage[toc,title,page]{appendix}
\usepackage[hidelinks]{hyperref}
\usepackage{url}

\hypersetup{
    colorlinks=true,
    linkcolor=green,
    citecolor=blue,
    filecolor=magenta,
    urlcolor=cyan,
    pdftitle={Sharelatex Example},
    pdfpagemode=FullScreen,
}

\fancypagestyle{plain}{\chead{\texttt{\textcolor{brown}{Communications of BAO, Vol. ?, Issue ?, 20??, pp. ???-???}}}}
\title{\textbf{The synchrotron mechanism and the high energy flair from PKS 1510-089}}
\author[1]{Osmanov Z.N.\thanks{z.osmanov@freeuni.eduge, Corresponding author}}

\affil[1]{\scriptsize School of Physics, Free University of Tbilisi, 0183, Tbilisi,
Georgia}
\affil[2]{\scriptsize E. Kharadze Georgian National Astrophysical Observatory, Abastumani, 0301, Georgia}

\begin{document}
\pagestyle{empty}
\newpage
\pagestyle{fancy}
\label{firstpage}
\date{}
\maketitle

\begin{abstract}
In order to understand the role of the synchrotron emission in the
high energy gamma-ray flares from PKS 1510-089, we study generation
of the synchrotron emission by means of the feedback of cyclotron
waves on the particle distribution via the diffusion process. The
cyclotron resonance causes the diffusion of particles along and
across the magnetic field lines. This process is described by the
quasi-linear diffusion (QLD) that leads to the increase of pitch
angles and generation of the synchrotron emission. We study the
kinetic equation which defines the distribution of emitting
particles. The redistribution is conditioned by two major factors,
QLD and the dissipation process, that is caused by synchrotron
reaction force. The QLD increases pitch angles, whereas the
synchrotron force resists this process. The balance between these
two forces guarantees the maintenance of the pitch angles and the
corresponding synchrotron emission process. The model is analyzed
for a wide range of physical parameters and it is shown that the
mechanism of QLD provides the generation of high energy (HE)
emission in the GeV energy domain. According to the model the lower
energy, associated with the cyclotron modes, provokes the
synchrotron radiation in the higher energy band.
\end{abstract}

\section{Introduction }
During the last decade the observations in the high and the very
high energy domains have stimulated theoretical studies of the
corresponding astrophysical objects. Recently the AGILE
(Astro-rivelatore Gamma a Immagini LEggero) collaboration announced
the discovery of the high energy (HE) $\gamma$-rays of PKS 1510-089
\citep{agilebl}, which is a nearby blazar ($z=0.361$)
belonging to the class of the flat radio quasars. The observations
on the mentioned source was performed in the period $9-30$ March
$2009$. According to the observational data, PKS 1510-089 showed
extreme variability in $\gamma$-rays. It was found that the HE
photon spectrum can be well described by the photon index $\Gamma =
1.95\pm 0.15$, the following average integral flux above $100$MeV,
$(311\pm 21)\times 10^{-8}$photons cm$^{-2}$s$^{-1}$, and a peak
integral flux $(702\pm 131)\times 10^{-8}$photons cm$^{-2}$s$^{-1}$
detected on 25 of March \citep{agilebl}. The monitoring
of the blazar between September 2008 and June 2009 was performed by
the Fermi-LAT telescope and a complex variability at optical, UV and
$\gamma$-ray bands was found \citep{fermibl}. During
this period three $\gamma$-ray flares have been detected, with the
brightest isotropic luminosity $2\times 10^{48}$erg/s on 2009 March
26. It has been found that the flux for energies above $200$MeV
reaches its peak value $2.4\times 10^{-7}$ photons $^{-2}$s$^{-1}$.

Obviously it is assumed that the HE emission is produced by the
inverse Compton mechanism \citep{blan} or curvature
radiation \citep{g96,tg05}. The synchrotron process is
supposed to be responsible only for producing relatively low energy
photons, because in strong magnetic fields the cooling timescale is
small compared with the kinematic timescale, that leads to the
transition of relativistic electrons to their ground Landau states,
resulting in the damping of the emission process.

Unlike the standard mechanism explaining the HE radiation, we apply
the so-called quasi-linear diffusion (QLD) that prevents the pitch
angles from damping, sustaining the synchrotron process. In
particular, it is believed that plasmas in pulsar magnetospheres
with strong magnetic fields may induce the unstable cyclotron waves
\citep{kmm}. These unstable modes in turn feedback on
relativistic electrons and by means of the diffusion influence the
particle distribution along and across the magnetic field lines \citep{lomin,machus1}. Under certain conditions the physical
system reaches the balance between the dissipation factors and the
diffusion, leading to saturation of the pitch angles. Therefore,
despite the efficient synchrotron losses, the pitch angles are
maintained, which provides a continuous emission process.

The mechanism of QLD was successfully applied to pulsars
\citep{malmach,nino,difus,difus1,ninoz}, anomalous pulsars \citep{comb1}, the black hole located in the center of  Milky Way, SgrA$^{\star}$ \citep{sgr} and active
galactic nuclei (AGN) \citep{difus3,difus4,difus6}. In
these papers it is shown that the QLD might provide the simultaneous
generation of relatively low frequency waves and the HE
$\gamma$-rays. On the other hand, it is observationally evident that
some AGN reveal the strong correlation of HE and low frequency
emission \cite{bloom,giro}, therefore the role of the
QLD might be important for these sources. In particular, in Ref.
\cite{difus3} the excitation of $X$-rays connected to the radio
waves has been studied. The similar method was developed to examine
the possibility of strong correlation of HE emission and
submillimeter/infrared radiation \citep{difus4} and
study the generation of high and very high energy $\gamma$-rays
strongly connected to radio emission \citep{difus6}.

Unlike the aforementioned articles, in the present work we focus on
the concrete AGN. According to the observational data presented by
AGILE collaboration \citep{agilebl}, the HE flare was
detected during 9-30 March 2009 from blazar PKS 1510-089. We apply
the mechanism of the QLD to the mentioned AGN and analyze the
synchrotron radiation processes in producing the observed HE
($>100$MeV) photons. According to the standard theory of the
synchrotron emission, it is assumed that the magnetic field is
chaotic along the line of sight, therefore, in the framework of this
approach the pitch angles vary in the broad interval, $0,\pi$ \citep{ginz}. Contrary to this scenario, the QLD prevents
the pitch angles from damping, restricting their values.

The paper is arranged in the following way. In section~II, we
introduce the theory of the QLD. In section~III, we apply the model
to the blazar PKS 1510-089 and in section~IV, we summarize our
results.

\section[]{Main Consideration}
In this section we present our model and apply it to BL Lac PKS
0548-322. This is a low redshift blazar with the central
supermassive black hole (SMBH) having the following mass $M\approx
5.4\times 10^8M_{\odot}$ \citep{fermibl}, where
$M_{\odot}\approx 2\times 10^{33}$g is the solar mass.  Around the
SMBH, in a region of the lengthscale, $l\sim 10^{14-15}$cm, the
magnetic field is strong enough to provide the frozen-in condition.
In the magnetospheres of AGNs, Lorentz factors of the magnetospheric
plasma particles lie in a broad interval, ranging from $\sim 1$ to
$10^8$ \citep{osm7,ra08}. For simplicity we consider
two component plasmas: the relatively low energy electron-positron
plasma component ($\gamma_p$) and the highly relativistic electrons
- the so-called beam component ($\gamma_b$). As we have already
mentioned, the relativistic particles will undergo strong
synchrotron losses. In particular, the synchrotron cooling timescale
for the beam electrons is given by the following expression
$t_{syn}\sim\gamma_b mc^2/P_{syn}$, where $m$ is the electron's
mass, $c$ is the speed of light, $P_{syn}\approx
2e^4\gamma_b^2B^2/3m^2c^3$ is the single particle synchrotron
emission power, $e$ is the electron's charge and $B$ is the magnetic
induction. By considering the relativistic electrons one can show
that the synchrotron cooling timescale, $t_{cool}$, is of the order
of $0.05\times\gamma_{b8}^{-2}B_{10}^{-2}$s
($\gamma_{b8}\equiv\gamma_b/10^8$, $B_{10}\equiv B/10G$), whereas
the kinematic timescale, $t_{kin}\sim l/c\approx 3000\times l_{14}$s
($l_{14}\equiv l/10^{14}cm$), for reasonable parameters is much
smaller than $t_{syn}$. Moreover, the condition $t_{cool}/t_{syn}\ll
1$ becomes more strict for higher luminosity sources, or during
$\gamma$-ray flares, since the equipartition magnetic field becomes
higher in these cases. Therefore, due to the strong magnetic field,
the synchrotron emission in the magnetosphere of PKS 1510-089 is
strongly suppressed by the energy losses and the particles rapidly
transit to their ground Landau states and generation of radiation is
stopped.

The situation drastically changes by means of the excited cyclotron
waves. This problem was considered for pulsars by 
\cite{kmm}. In particular, since the magnetic field
near the pulsar's surface is very strong, any transverse momenta of
relativistic electrons are lost and the corresponding distribution
function of electrons from the very beginning of motion becomes
one-dimensional and anisotropic. \cite{kmm} showed that for the aforementioned conditions the anomalous
Doppler effect leads to the generation of the pure transversal
cyclotron modes
\begin{equation}\label{omt}
\omega_t \approx kc\left(1-\delta\right),\;\;\;\;\; \delta =
\frac{\omega_p^2}{4\omega_B^2\gamma_p^3},
\end{equation}
where $k$ is the modulus of the wave vector, $\omega_p \equiv
\sqrt{4\pi n_pe^2/m}$ is the plasma frequency, $n_p$ is the plasma
number density and $\omega_B\equiv eB/mc$ is the cyclotron
frequency. We assume that energy in plasmas is distributed
uniformly, $n_p\approx n_b\gamma_b/\gamma_p$ ($n_b$ is the number
density of the beam electrons). As it was shown in \citep{machus1} the mentioned cyclotron wave is characterized by
the following frequency
\begin{equation}\label{om1}
\nu\approx \frac{\omega_B}{2\pi\delta\cdot\gamma_b}.
\end{equation}

The major requirement for excitation of the cyclotron modes is
considerably strong magnetic field, so that the magnetic energy
density, $W_B$, can exceed the plasma energy density, $W_p$. One can
straightforwardly show that for the light cylinder lengthscales the
mentioned condition $W_B/W_p>1$ writes as
\begin{equation}\label{ww}
   \frac{W_B}{W_p}\approx 100\times \frac{10^8}{\gamma_b}\times \frac{100
   cm^{-3}}{n_b}>1.
\end{equation}
This in turn means that the maximum value of the beam number
density, allowing the excitation of the cyclotron waves is of the
order of $2\times 10^4$cm$^{-3}$. If this condition is satisfied,
the unstable cyclotron waves are excited, which by means of the
feedback, through the diffusion process, affect the distribution of
particles, creating the pitch angles.

On the other hand, since the emitting particles are extremely
energetic they undergo the synchrotron radiation reaction force \citep{landau}
\begin{equation}\label{f}
    F_{\perp}=-\alpha\frac{p_{\perp}}{p_{\parallel}}\left(1+\frac{p_{\perp}^{2}}{m^{2}c^{2}}\right),
    F_{\parallel}=-\frac{\alpha}{m^{2}c^{2}}p_{\perp}^{2}.
\end{equation}
Unlike the role of the diffusion, this force is responsible for the
dissipation process, decreasing the pitch angle. Here
$\alpha=2e^{2}\omega_{B}^{2}/3c^{2}$ and $p_{\perp}$ and
$p_{\parallel}$ are transversal and longitudinal components of
momentum respectively. The corresponding kinetic equation governing
the mentioned mechanism is given in a series of works
\citep{ninoz}
\begin{eqnarray}\label{kin1}
\frac{\partial\textit{f }^{0}\left(\mathbf{p}\right)}{\partial
    t}+\frac{\partial}{\partial
p_{\parallel}}\left\{F_{\parallel}\textit{f
}^{0}\left(\mathbf{p}\right)\right\}+\frac{1}{p_{\perp}}\frac{\partial}{\partial
p_{\perp}}\left\{p_{\perp}F_{\perp}\textit{f
}^{0}\left(\mathbf{p}\right)\right\}=\nonumber
\\=\frac{1}{p_{\perp}}\frac{\partial}{\partial p_{\perp}}\left\{p_{\perp}\left(D_{\perp,\perp}\frac{\partial}{\partial p_{\perp}}+D_{\perp,\parallel}\frac{\partial}{\partial
p_{\parallel}}\right)\textit{f
}^{0}\left(\mathbf{p}\right)\right\}+\nonumber
 \\
+\frac{\partial}{\partial
p_{\parallel}}\left\{\left(D_{\parallel,\perp}\frac{\partial}{\partial
p_{\perp}}+D_{\parallel,\parallel}\frac{\partial}{\partial
p_{\parallel}}\right)\textit{f }^{0}\left(\mathbf{p}\right)\right\},
\end{eqnarray}
where $\textit{f }^{0}\left(\mathbf{p}\right)$ is the distribution
function,
\begin{eqnarray}\label{dkoef}
     \left(%
\begin{array}{c}
  D_{\perp,\perp} \\
  D_{\perp, \parallel}=D_{\parallel\perp} \\
  D_{\parallel,\parallel} \\
\end{array}%
\right)=\left(%
\begin{array}{c}
  D\delta|E_{k}|^{2} \\
  -D \psi|E_{k}|^{2} \\
  D \psi^{2}\frac{1}{\delta}|E_{k}|^{2} \\
\end{array}%
\right),
\end{eqnarray}
are the diffusion coefficients \citep{ninoz}, $E_{k}$
is the electric field, square of which is given by $|E_k|^2 =
mc^3n_b\gamma_b/(4\pi\nu)$ \citep{malmach}, $\psi =
p_{\parallel}/p_{\perp}$ is the pitch angle and $D=e^{2}/8c$.
Normally the pitch angles are very small, $\psi\ll 1$, therefore,
one can reduce Eq. (\ref{kin1}) by using the relation
$\partial/\partial p_{\perp}>>\partial/\partial p_{\parallel}$
\citep{ninoz}

In the framework of the model the pitch angles saturate due to the
balance between the diffusion and the dissipation factors. This
leads to the stationary regime ($\partial/\partial t = 0$) and the
solution of Eq. (\ref{kin2}) writes as
\begin{equation}\label{ff}
    \textit{f}(p_{\perp})=C exp\left(\int \frac{F_{\perp}}{D_{\perp,\perp}}dp_{\perp}\right)=Ce^{-\left(\frac{p_{\perp}}{p_{\perp_{0}}}\right)^{4}},
\end{equation}
where
\begin{equation}\label{pp0}
     p_{\perp_{0}}=\left(\frac{4\gamma_bm^3c^3D_{\perp,\perp}}{\alpha}\right)^{1/4}.
\end{equation}

\begin{eqnarray} \label{kin2}
    \frac{\partial\textit{f }^{0}}{\partial
    t}+\frac{1}{p_{\perp}}\frac{\partial}{\partial p_{\perp}}\left(p_{\perp}
    F_{\perp}\textit{f }^{0}\right)
    =\frac{1}{p_{\perp}}\frac{\partial}{\partial p_{\perp}}\left(p_{\perp}
D_{\perp,\perp}\frac{\partial\textit{f }^{0}}{\partial
p_{\perp}}\right).
\end{eqnarray}
\begin{figure}
  \resizebox{\hsize}{!}{\includegraphics[angle=0]{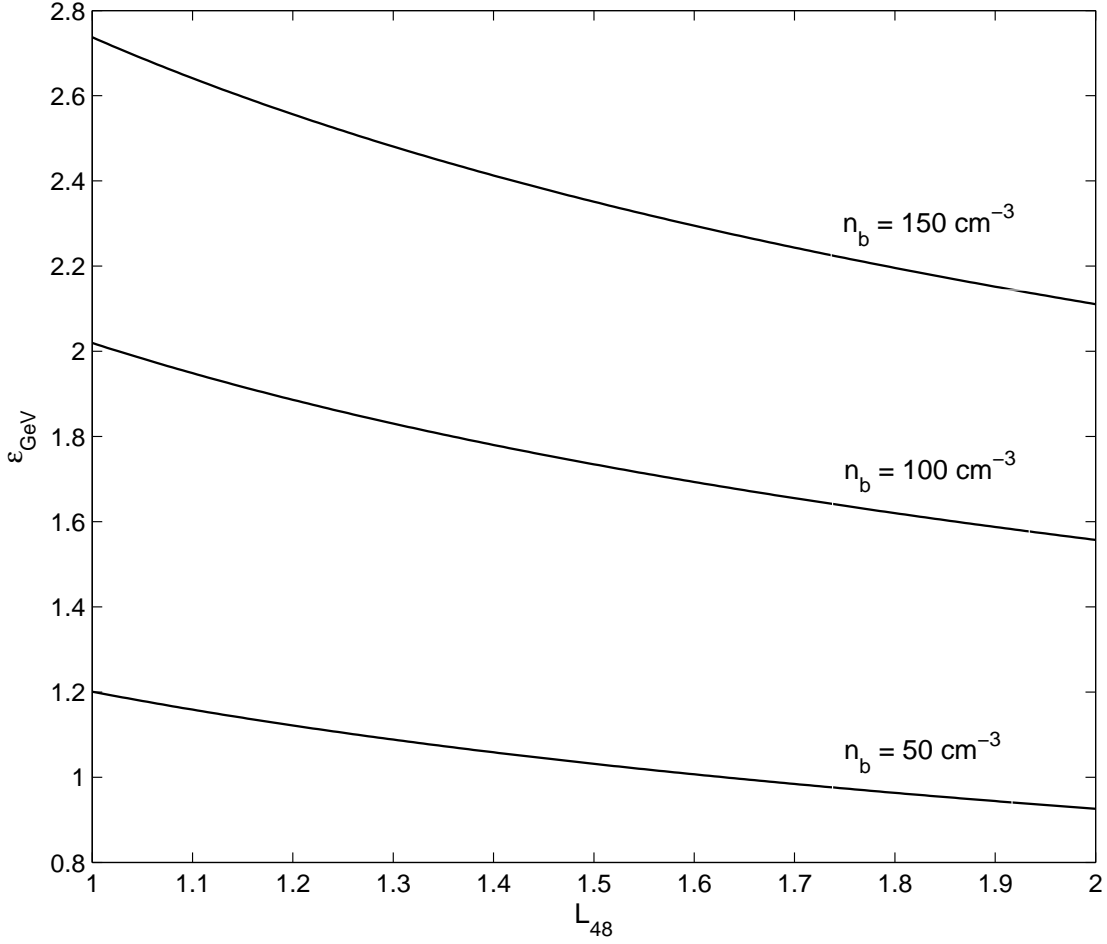}}
  \caption{Behaviour of $\epsilon_{GeV}$ with respect to the dimensionless
  luminosity $L_{48}$.
The set of parameters is: $M\approx 5.4\times 10^8M_{\odot}$, $r =
70R_g$, $n_b\in [50;100;150]$ cm$^{-3}$, $\gamma_p\approx 1$ and
$\gamma_b = 2\times 10^8$.}\label{fig1}
\end{figure}

It is evident from Eq. (\ref{ff}) that only a tiny fraction of
electrons have the transversal momentum sufficiently exceeding the
value of $p_{\perp_{0}}$. Therefore, it is worth estimating the
average value of $p_{\perp}$
\begin{equation}\label{avp}
\langle p_{\perp}\rangle
 = \frac{\int_{0}^{\infty}p_{\perp} \textit{f}(p_{\perp})dp_{\perp}}
 {\int_{0}^{\infty}\textit{f}(p_{\perp})dp_{\perp}}
\approx \frac{p_{\perp_{0}}}{2},
\end{equation}

which naturally describes the mean value of the pitch angles,
$\langle\psi\rangle = \langle p_{\perp}\rangle/p_{\parallel}$, and
the corresponding photon energy produced by the synchrotron emission
\citep{Lightman}
\begin{equation}\label{eps}
\epsilon_{_{MeV}}\approx 2.5\times 10^{-15}\frac{\gamma_b
p_{\perp_{0}}B}{mc}.
\end{equation}
Generally speaking, strong magnetic field leads to efficient energy
losses, resulting in one-dimensional distribution function. This in
turn, creates all necessary conditions for excitation of the
cyclotron waves, leading to the final result of this complex
process: the creation of the pitch angles, that inevitably causes
synchrotron radiation. Therefore, as it is clear from the
consideration, the strong synchrotron energy losses do not impose
any constraints.

\begin{figure}
  \resizebox{\hsize}{!}{\includegraphics[angle=0]{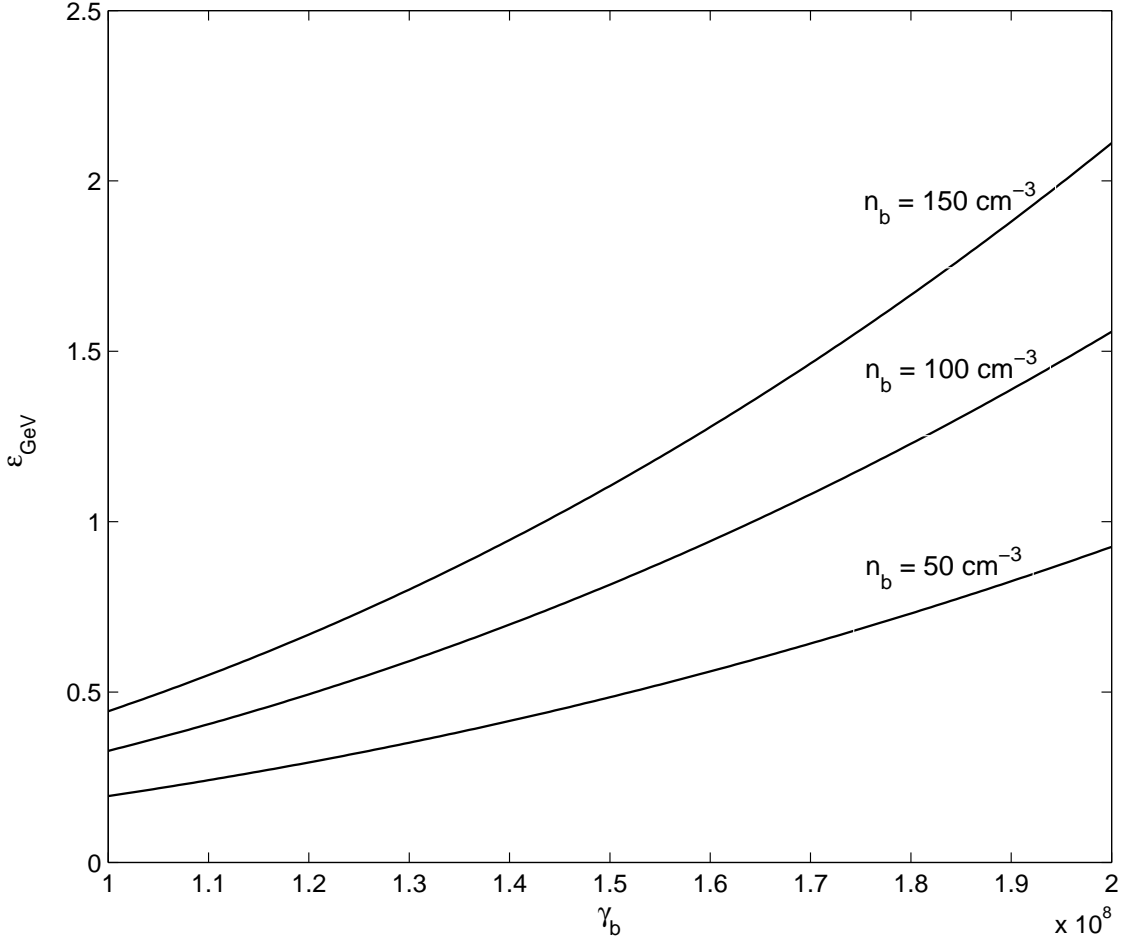}}
  \caption{Behaviour of $\epsilon_{GeV}$ with respect to the
  Lorentz factor of relativistic electrons.
The set of parameters is: $M\approx 5.4\times 10^8M_{\odot}$, $r =
70R_g$, $n_b\in [50;100;150]$ cm$^{-3}$, $\gamma_p\approx 1$, $L =
2\times 10^{48}$erg/s.}\label{fig2}
\end{figure}

\section{Discussion}

In this section we study properties of the HE emission of the blazar
PKS 1510-089. According to \cite{fermibl}, between September
2008 and June 2009 the Fermi LAT has detected three $\gamma$-ray
flares. The most luminous isotropic flare with the estimated
luminosity $7\times 10^{47}$erg/s was detected during 2009-03-10 and
2009-04-09, with the brightest daily luminosity of the order of
$2\times 10^{48}$erg/s (March 29 2009). The equipartition magnetic
field depends on the source luminosity, $B\approx\sqrt{2L/(r^2c)}$
($r$ is the distance from blazar). Therefore, effectiveness of the
QLD strongly depends on the flare efficiency, since the diffusion
coefficient depends on $B$ (see Eq. {\ref{dkoef}}).

In Fig. \ref{fig1} we show the dependence of $\epsilon_{GeV}$ on the
dimensionless luminosity, $L_{48}\equiv \frac{L}{10^{48}erg/s}$ for
different densities of the beam electrons. The set of parameters is:
$M\approx 5.4\times 10^8M_{\odot}$, $r = 70R_g$, $n_b\in
[50;100;150]$ cm$^{-3}$, $\gamma_p\approx 1$ and $\gamma_b = 2\times
10^8$. As it is evident from the figure, the photon energy is a
continuously decreasing function of the source luminosity. This is a
direct consequence of the following behaviour $p_{\perp_{0}}\propto
L^{-5/8}$ (see Eq. \ref{dkoef}), which means that for more luminous
sources the pitch angles, $\langle\psi\rangle\propto p_{\perp_{0}}$,
and the corresponding synchrotron photon energies will be lower. As
it is clear from the plots, by increasing the density, the
corresponding photon energy increases as well. In particular, by
taking into account Eqs. (\ref{pp0}-\ref{eps}) one can see that
$\epsilon_{_{MeV}}\propto n_b^{1/2}$. From the plots we see that the
QLD may provide HE emission in the GeV energy domain if the beam
electrons' Lorentz factors and number density are of the order of
$2\times 10^8$ and $50;100;150$cm$^{-3}$, respectively and
$\gamma_p\approx 1$.

In Fig. \ref{fig2} we show the dependence of the photon energy on
the Lorentz factors of the beam electrons. The set of parameters is
the same as for Fig. \ref{fig1}, except $L = 2\times 10^{48}$erg/s
and $\gamma_b\in [1-2]\times 10^8$. As we see from the figures, the
QLD may provide the synchrotron emission from multi MeV to several
GeV, which is in a good agreement with the observations \citep{agilebl,fermibl}. On the other hand, it is worth noting
that according to the multifrequency campaigns, the spectral energy
distribution (SED) of the blazar PKS 1510-089 shows two clear peaks,
one in the GeV energy domain and another peak near $10^{13}$Hz. This
particular peak is formed by mildly relativistic electrons in a
different location, where magnetic induction is supposed to be of
\begin{figure}
  \resizebox{\hsize}{!}{\includegraphics[angle=0]{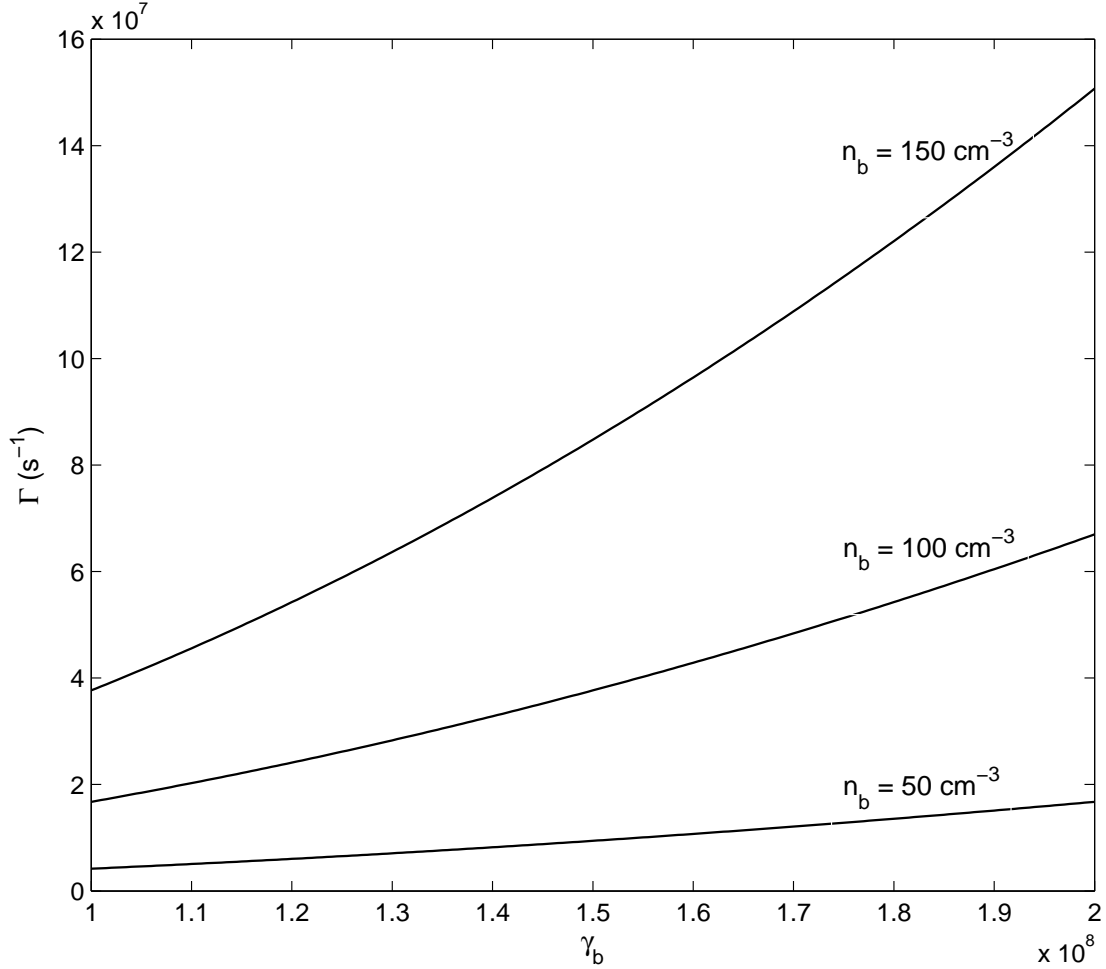}}
  \caption{Behaviour of $\Gamma$ with respect to the
  Lorentz factor of relativistic electrons.
The set of parameters is: $M\approx 5.4\times 10^8M_{\odot}$, $r =
70R_g$, $n_b\in [50;100;150]$ cm$^{-3}$, $\gamma_p\approx 1$, $L =
2\times 10^{48}$erg/s.}\label{fig3}
\end{figure}the order of $3$G. In particular, by assuming that the corresponding
emission is produced by the synchrotron mechanism, one can
straightforwardly show from $\nu_{1}\approx 2.9\times 10^6
\gamma^2B$Hz \citep{Lightman} that the mildly
relativistic electrons with $\gamma\sim 50$ can explain the first
peak of SED. On the other hand, it is clear that energy losses in
this case are not efficient, because $t_{syn}\sim 10^6$s exceeds the
kinematic timescale and hence the synchrotron process does not
require any additional mechanism for maintaining the radiation.
Therefore, according to our model, the HE emission comes from the
nearby zone of the light cylinder surface $r\sim 5.6\times
10^{15}$cm (inner magnetosphere), whereas the first peak with
$10^{13}$Hz is formed in an outer magnetosphere of PKS 1510-089,
where the magnetic field is lower.

Since the mechanism of the QLD is driven by the unstable cyclotron
waves, it is important to investigate the effectiveness of the
instability and the corresponding growth rate. \cite{kmm}
has shown that the increment of the cyclotron instability is
given by
\begin{equation}\label{inc1}
\Gamma = \frac{\omega_b^2}{2\nu\gamma_p},
\end{equation}
where $\omega_b\equiv\sqrt{4\pi n_b e^2/m}$ is the plasma frequency
of beam electrons. In Fig. \ref{fig3} we show the behaviour $\Gamma
(\gamma_b)$. The list of parameters is the same as for Fig.
\ref{fig2}. As it is clear from the plots, depending on the physical
parameters, the Growth rate varies in a broad interval $[5\times
10^6-1.5\times 10^8]$s$^{-1}$, leading to the following timescales
$t_{ins}\sim 1/\Gamma\sim [7\times 10^{-9}-2\times 10^{-7}]$s. On
the other hand, the kinematic timescale $t_{kin}\sim r/c$ is of the
order of $2\times 10^5$s, which exceeds that of the instability by
many orders of magnitude. Therefore, the cyclotron instability is
extremely efficient to provide the excitation of the cyclotron waves
and maintain the synchrotron emission regime.

\section{Summary}
\begin{enumerate}

      \item We study the recently detected (by AGILE and Fermi-LAT)
      HE $\gamma$-ray emission from the blazar PKS 1510-089. Applying the
      mechanism of the QLD we argue that in spite of
      the strong synchrotron energy losses, the diffusion is
      effective enough to provide the observed energies of the
      source.

      \item It is shown that the excited unstable cyclotron waves
      strongly influence the distribution function of relativistic
      electrons by means of the quasi-linear diffusion.
      This in turn maintains the pitch angles from
      damping and provides the continuous synchrotron emission
      process.

      \item We found that under favorable conditions the
      QLD may guarantee the HE emission with energies from multi MeV to several GeV,
      that is in a good agreement with the observations of AGILE and
      Fermi-LAT.

      \end{enumerate}

As we see, the synchrotron mechanism may guarantee emission in the
HE domain. Another important issue that we would like to address is
the emission spectrum. As we have already mentioned in the
introduction, according to the standard theory of the synchrotron
radiation \citep{ginz} the magnetic field is chaotic
along the line of sight, therefore, the corresponding values of the
pitch angles lie in a broad interval, $\psi\in (0;\pi)$. Unlike this
case, in the framework of the QLD the pitch angles are restricted by
the dissipation factors leading to a certain spectral picture.
Therefore, sooner or later we are going to examine this particular
problem as well.

\section*{Acknowledgments}

The author is grateful to Prof. George Machabeli for valuable discussions
and helpful suggestions.

\end{document}